\documentclass{moriond}

\def\be{\begin{equation}}
\def\ee{\end{equation}}
\def\bea{\begin{eqnarray}}
\def\eea{\end{eqnarray}}

\newcommand{\Photo}{\includegraphics[height=35mm]{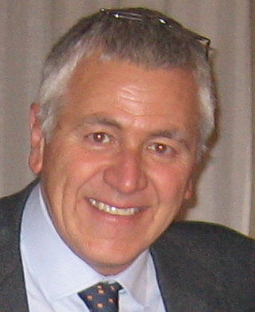}}

\begin{document}
\vspace*{4cm}
\title{Analysis and discussion of the recent $W$ mass measurements}

\author{ GIGI ROLANDI }

\address{CERN Ginevra and Scuola Normale Superiore  Pisa}

\maketitle\abstracts{
The ATLAS and CDF measurements of the W mass are compared discussing some similarities and differences in the quoted systematic uncertainties.}

\section{Introduction}

The Organizing Committee asked me to prepare a short presentation to stimulate the discussion on the recent $W$ mass measurement~\cite{ATLAS1}
of the ATLAS collaboration nicely presented by Nansi Andari~\cite{ATLAS2}. I decided to show a comparison of this
measurement with the {\it old} CDF measurement~\cite{CDF1} that has the same precision of the ATLAS one: 19 MeV. 
Since I refer the reader quite often to figures published in the two papers it is convenient to read this note having the two papers~\cite{ATLAS1,CDF1} at hand.\\

Both measurements are done using electronic and muonic decays of the W. The CDF sample consists of  1.1 Millions $W$,
 evenly distributed between $W^+$ and $W^-$, produced in $p\bar{p}$ collisions at $\sqrt{s}=2 \;{\rm TeV}$ ; the ATLAS sample comprises 7.8 Millions $W^+$ and
 5.9 Millions $W^-$ produced in $pp$ collisions at $\sqrt{s}=7\;  {\rm TeV}$.\\
 
The $W$ mass is measured in $W\rightarrow \ell \nu$ decays from the transverse momentum distribution of the charged lepton\footnote {In the following I will use the word lepton to indicate the charged lepton}   ($p_T$)  and from the transverse mass distribution
 ($M_T$), where the information of transverse momentum of the lepton is combined with the transverse momentum of the hadronic system recoiling to the W  ($\vec{h}$): $M_T^2=2\left( p_T|\vec{p_T}+\vec{h}|+p_T^2+\vec{p_T}\vec{h}\right)$. The CDF collaboration uses also the missing transverse energy distribution $p_T^{\nu}=|\vec{p_T}+\vec{h}|$. For each experimental distribution ($p_T$,$M_T$,$p_T^{\nu}$) a value of the W mass is determined comparing the measured distribution to templates generated with different W mass hypotheses using a carefully calibrated MonteCarlo. The correlated values of the W mass are combined in order to minimize the total uncertainty.\\
 
 In both analyses the events are selected requiring one identified lepton matched to the trigger and   $30\,{\rm GeV}<p_T$ , $30\,{\rm GeV}<p_T^{\nu}$ and $60\,{\rm GeV}<M_T$. The cut of the recoil is $h<15$ GeV in CDF and $h<30$ GeV in ATLAS. The comparison between the measured distribution and the template distributions is done in a limited region (fit range) of the variable of interest. 
 The fit ranges in CDF are  $32<p_T<48$ GeV and $65<M_T<90$ GeV in CDF and $32<p_T<45$ GeV and $66<M_T<99$ GeV in ATLAS.\\
  
 The result of the CDF analysis is $M_W=80\,387\pm12({\rm stat.})\pm 15({\rm syst.})$ MeV; the result of the ATLAS analysis is $M_W=80\,370\pm 7({\rm stat.})\pm 11({\rm exp.\, syst.})\pm 14({\rm mod.\, syst.})$ MeV. The total uncertainty is 19 MeV in both analyses. The weights of the muon and electron samples are similar in the two analyses : 0.62/0.38 in CDF and 0.57/0.43 in ATLAS. The weights of the $M_T$ / $p_T$ fits in ATLAS are 0.14/0.86; the weights of the $M_T$/ $p_T$ / $p_T^{\nu}$ fits in CDF are 0.53/0.31/0.16. The ATLAS measurement is mainly done using the $p_T$ observable while the CDF one is mainly done using the $M_T$ observable that is less sensitive to prior knowledge of the transverse momentum distribution of the $W$.\\ 
 
 Table~\ref{TAB1} shows a breakdown of the uncertainties in the two analyses using separately the $p_T$ and $M_T$ observables. I have followed the uncertainties breakdown structure presented in the CDF paper and I have adapted the uncertainties quoted in the ATLAS paper to this somewhat different list.
 
\begin{table}
\caption[]{Break down of the uncertainties in the $W$ mass measurement in CDF and ATLAS analyses. All figures in MeV}
\label{TAB1}
\centering
\begin{tabular}{lcccc}
\hline
& CDF/$p_T$&CDF/$M_T$&ATLAS/$p_T$&ATLAS/$M_T$\\
\hline
Statistical &16&15&7.2&9.6\\
Lepton Scale and Resolution &7&7&6.5&6.5\\
Recoil Scale and Resolution &5.5&6&2.5&13\\
Backgrounds &4&3.5&4.6&8.3\\
PDFs &9&10&9&10.2\\
W transverse momentum model / QCD &9&3&8.3&9.6\\
Photon Radiation/EWK &4&4&5.7&3.4\\
\hline
\end{tabular}
\end{table}
 \section{Statistical uncertainty}
 The ATLAS sample is about 12 times the CDF sample and one would naively expect a reduction of the statistical uncertainty by a a factor $\sqrt{12}\simeq 3.5$. Table~\ref{TAB1} shows
 ratios of 2.2 and 1.5 for the statistical uncertainties of the $p_T$ and $M_T$ analyses. The $p_T$ spectra in the two experiments are quite similar and the smaller fit range in ATLAS accounts for less than 20\%
 reduction of the statistics. The $M_T$ spectrum in ATLAS is some 20\% larger due to the worse recoil resolution. These effects do not explain the observed difference between the statistical uncertainties.\\
 
 The ATLAS sample is divided in categories that are also fitted separately. The sample of $W^+\rightarrow \mu\nu$ with $\eta<0.8$ has a  roughly 1.2 Million events as can be estimated from figure 20 of reference~\cite{ATLAS1}. The statistical uncertainty on the $p_T$ and $M_T$ fits done on this sample are 22 and 29 MeV (see table 10 of reference~\cite{ATLAS1}), a factor 1.3 and 1.8 larger than the CDF quoted statistical uncertainty.
 
 \section{Uncertainty due to the recoil}
 The systematic uncertainty in the $p_T$ fit due to the recoil is caused by the event selection that depends on the recoil~\cite{kotwal}. This uncertainty is evaluated in CDF modeling the recoil distribution using Z data and assessing the uncertainty on $M_w$ from the difference in the measured mass obtained repeating the $W$ mass fit with a 1$\sigma$ variation of the recoil model. In CDF this results in an uncertainty of about 6 MeV in the $p_T$ fit and is caused by events that statistically are accepted or rejected by the selection. The worse the recoil resolution the larger is this effect~\cite{kotwal} .\\
 
The measurement of the recoil in CDF has a response of 65\% and a resolution on the {\it true} recoil (eg corrected for the response)  of 7 GeV. In ATLAS the response is larger, close to 90\% and the resolution is about 10 GeV. The systematic uncertainty due to recoil on the $p_T$ fit in ATLAS is 2.5 MeV that is small compared to the 6 MeV of CDF.  I wonder if this is due to the optimization procedure of ATLAS that is choosing the fit range that minimizes the total uncertainty. In CDF the fit range is decided a priori.

\section{Uncertainty due to the PDFs}
 In CDF the uncertainty due to the Parton Distribution Functions (PDF) is computed repeating the $M_w$ fits with the different independent variations of the PDFs \footnote{ The PDF distributions are given as a central set of distributions for each relevant parton and a number of replicas each corresponding to  1$\sigma$ of the uncertainty of the PDF fit. These replicas correspond to independent variations in the PDF fit.} and assessing the uncertainty as
 the sum in quadrature of the variations of $M_w$ resulting from each fit. They use CTEQ6.6 and MSTW2008 PDFs. The result is a systematic uncertainty of 10 MeV. \\
 
 ATLAS uses CT10nnlo PDFs and has a procedure similar to the one adopted by CDF. In ATLAS the fit is done in categories dividing the sample in four lepton rapidity bins, separately for $W^+$ and $W^-$. The result of the fits shows large beneficial correlation when the PDF distributions are varied. The PDF uncertainty in the fits of $W^+$ and $W^-$ separate samples  is 13.5 and 14.5 MeV respectively for the $p_T$ fit and 16.9 and 16.2 MeV for the $M_T$ fit. The uncertainty on the fits combining $W^+$ and $W^-$  are 9 MeV and 10 MeV respectively , i.e. a reduction of 40\% of the uncertainty !  One of the reasons is that the $c\bar{s}$ contribution to the $W^+$ sample is almost identical (bin by bin in the variable of interest) to the $s\bar{c}$ contribution to the $W^-$. When combining the two samples the contribution to the PDF uncertainty of the less known $s$ and $c$ quarks parton distribution functions is reduced.\\
 
 Another interesting effect shown by the ATLAS analysis is that the PDF uncertainties in the different  $\eta$ categories have also beneficial correlations. As shown in table~\ref{TAB2} the PDF uncertainty is reduced when combining the different $\eta$ bins.
 
 \begin{table}
 \centering
 \label{TAB2}
 \caption []{Uncertainty due to Parton Distribution Functions in the fit done in different $\eta$ categories in the ATLAS analysis.  Figures from table 10 and table 11 of reference~\cite{ATLAS1}. All figures are in MeV.}
 \begin{tabular}{l|cccccc}
 \hline
 $\eta$ range & 0-0.8&0.8-1.4&1.4-2.0&2.0-2.4&~~~&combined\\
 \hline
 PDF $W^+$/$p_T$ fit & 24.7 & 20.6 & 25.2 & 31.8 &&14.5\\
 PDF $W^+$/$M_T$ fit& 28.4&23.3&27.2&32.8&&16.9\\
 \hline
 \end{tabular}
 \end{table}
 
 \section{Uncertainty due to the modeling of the W transverse momentum}
 The W transverse momentum (pt) distribution is a very important ingredient of the W mass analysis. It must be known to very good precision when fitting the lepton $p_T$ distribution 
 and its effect is somewhat mitigated in the fit of the $M_T$ distribution depending on the available statistics and on the response and resolution on the measurement of the recoil.\\
 
 The strategy adopted by ATLAS and CDF is to measure and model precisely the Z pt distribution and use the model to predict the W pt distribution.\\
 
 CDF fits the measured Z pt distribution using RESBOS~\cite{RESBOS} determining the  two parameter of interest ($g_2$ and $\alpha_s$) and gets an agreement at the 2\% level between the fitted function and the measured distributions as shown in figure 5 of reference~\cite{CDF1}. They use the same function to predict the W pt distribution and they quote an uncertainty on the $W$ mass of 9 and 3 MeV for the $p_T$ and $M_T$ fits computed from the variations in 1$\sigma$ of the fitted parameters of the model. The smaller uncertainty in the $M_T$ fit is what one expects because of the reduced dependence of the $W$ transverse mass on the boson pt distribution.\\
  
 ATLAS uses PYTIA8  to fit the Z pt distribution varying the intrinsic transverse momentum of the incoming partons, the value of $\alpha_s$ used for the QCD ISR, and the value of the ISR infrared cut-off. The resulting tune (AZ tune) reproduces the Z pt spectrum with an agreement at 1-2\% level. When exporting the model to the W pt distribution the variation of the AZ tune fit parameters produce a systematic uncertainty of about 3 MeV. In addition ATLAS consider systematic uncertainties due to other contribution to the W production model including the charm mass, the variation of the factorization scale of the parton shower with heavy-flavour decorrelation, the parton shower PDF uncertainty and the uncertainty of the angular coefficients. The resulting uncertainties on the $W$ mass in the $p_T$ and $M_T$ fits from these sources are quite similar as shown in table  3 of reference~\cite{ATLAS1}, resulting in an uncertainty of 8 MeV and 9 MeV on the $P_T$ and $M_T$ fits.\\
 
\begin{figure}
\begin{minipage}{0.48\linewidth}
\centerline{\includegraphics[width=0.9\linewidth]{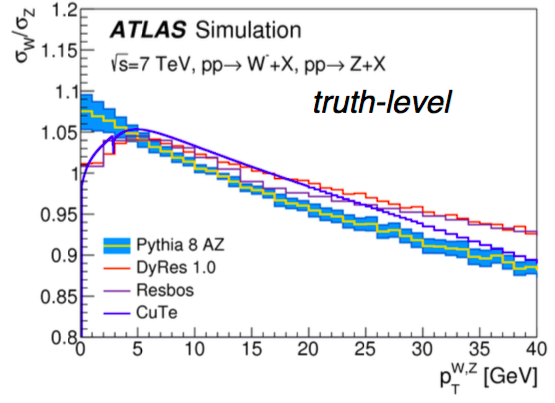}}
\end{minipage}
\hfill
\begin{minipage}{0.48\linewidth}
\centerline{\includegraphics[width=0.9\linewidth]{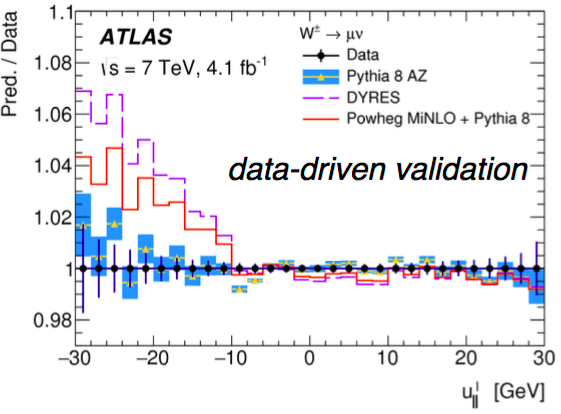}}
\end{minipage}
\hfill
\caption[]{Plots from the talk of N. Andari~\cite{ATLAS2}. Left: ratio between the Z and W pt spectra of different calculations/generators. The blue band shows the  systematic uncertainty in the ATLAS analysis. Right: $u_{||}$ is the component of the W recoil parallel to  direction of the lepton in W events. The plot shows the ratio between the predicted  and measured distributions}
\label{FIG1}
\end{figure}

 Figure~\ref{FIG1} left shows the ratio between the pt distribution of the W and Z bosons for the model used by the ATLAS analysis and other predictions of the boson pt spectrum. Resummed calculations like DYRES~\cite{DYRES} predict an harder spectrum compared to the one used in the analysis and the difference is not covered by the quoted systematic uncertainty band. In order to validate the model ATLAS measures in W events the recoil component parallel to the lepton direction $u_{||}$ (both vectors in the transverse plane). Due to events selection cuts, when $u_{||}$ is close to 30 GeV in absolute value the recoil is forced to be parallel or antiparallel to the lepton direction. The rate of goes to zero approaching  $u_{||}$  of 30 GeV from below and -30 GeV from above as can be seen in figure 19 of reference~\cite{ATLAS1}. Events where the W pt is parallel to the lepton appear near -30 GeV in  figure~\ref{FIG1} b. These are events where the lepton transverse momentum exceeds 60 GeV. The events near +30 GeV in the same plot are events were the W pt is antiparallel  to the lepton direction: the {\it neutrino} momentum exceeds 60 GeV  and the lepton momentum is larger than 30 GeV.  Because of the resolution and of the response of the recoil, the rate of events near $u_{||}\sim +30$ GeV  is larger than the rate near -30 GeV. The plot shows that near  $u_{||}\sim -30$ GeV the data are not in agreement with the resummed calculations that predict a larger cross section than what measured in data by some 5\%. This region is characterized by leptons with large transverse momentum that are produced in events where the boson pt is large. This is not the case in  the region near  $u_{||}\sim +30$ GeV. Here  the boson pt can also be quite small because of the resolution in the measurement of the recoil.\\

The question why resummed calculations do describe the Z pt distribution in $p\bar{p}$ at 2 TeV and do not describe the $pp$ data at 7 TeV remains open.

\section{Conclusions}
ATLAS has produced the first W mass measurement at LHC with the data collected at $\sqrt{s}=$ 7 TeV. This is a very welcome event and opens the road to future measurements using even larger data samples. This analysis shows that the systematic uncertainty due to PDFs can be kept under control and possibly brought below the 10 MeV level. There are still few aspects to be understood among which why the measured  Z pt distribution is not in agreement with  prediction using  resummation calculations.

\section*{Acknowledgments}

I would like to thank Patrick Janot and the organizing committee for inviting me to give  this presentation. 
Moriond Electroweak is in my opinion the most stimulating conference in HEP and it is always a pleasure
to participate actively to the works.

\end{document}